# Control of coherent backscattering by breaking optical reciprocity

Y. Bromberg, B. Redding, S. M. Popoff[†] and H. Cao*

*Department of Applied Physics, Yale University, New Haven, Connecticut 06520 USA*

[†] Current address:  *CNRS - LTCI Telecom ParisTech, 46 rue Barrault, 75013 Paris, France*

* hui.cao@yale.edu

**Abstract**

Reciprocity is a universal principle that has a profound impact on many areas of physics. A fundamental phenomenon in condensed-matter physics, optical physics and acoustics, arising from reciprocity, is the constructive interference of quantum or classical waves which propagate along time-reversed paths in disordered media, leading to, for example, weak localization and metal-insulator transition. Previous studies have shown that such coherent effects are suppressed when reciprocity is broken[1–3]. Here we show that by breaking reciprocity in a controlled manner, we can tune, rather than simply suppress, these phenomena. In particular, we manipulate coherent backscattering of light[4–7], also known as weak localization. By utilizing a non-reciprocal magneto-optical effect, we control the interference between time-reversed paths inside a multimode fiber with strong mode mixing, and realize a continuous transition from the well-known peak to a dip in the backscattered intensity. Our results may open new possibilities for coherent control of classical and quantum waves in complex systems.

**Introduction**

The reciprocity principle demands that waves which propagate along time-reversed paths exhibit the exact same transmission, no matter how complex the paths are[8]. It has profound and often surprising implications regarding the transport of classical and quantum waves in complex systems, as it poses a symmetry that is not distorted by disorder. When pairs of time-reversed paths interfere, reciprocity guarantees that the interference is constructive for any realization of disorder (Fig. 1(a)). This robust interference is the underlying mechanism of weak localization, a coherent correction to incoherent transport models like the Boltzman-Drude for electrical conductance and the radiative transfer equation for light[9]. The discovery of weak localization, originally for mesoscopic transport of electrons, marked a milestone in the research of complex coherent phenomena[10]. It established the importance of interference effects even when waves are randomly scattered, eventually leading to the celebrated strong (Anderson) localization. In optics, weak localization is manifested by an enhancement by a factor of 2 of the backscattered intensity, an effect coined coherent backscattering (CBS)[4–7].

The role of reciprocity in multiple scattering phenomena is elucidated when it is broken. For electrons, reciprocity is broken by magnetic fields that induce Ahraonov-Bohm oscillations recorded by magnetoresistance measurements[11,12]. In optics, previous studies on broken reciprocity with magneto-optical effects[1] or nonlinearity[2] in scattering systems showed suppression of the CBS enhancement: since time-reversed paths are no longer guaranteed to interfere constructively, the CBS peak disappears[3,13]. Here, we demonstrate a novel scheme for non-reciprocal tuning of the relative phase between all the pairs of time-reversed paths, which enables a continuous transition from the CBS peak to a dip. This work underlies a new class of control mechanism for complex coherent phenomena with broken reciprocity.

Since optical reciprocity and specifically CBS are universal phenomena, they can be studied in a wide range of scattering systems, such as paints, colloids, and biological tissue. Here we study for the first time CBS in multimode optical fibers with strong mode coupling. In recent years there has been an increasing interest in exploiting the high capacity of multimode fibers for numerous applications, including optical communication[14], imaging[15,16], and spectroscopy[17]. In this work, we show that multimode fibers also provide an excellent platform for studying fundamental aspects of mesoscopic transport, and in particular for controlling CBS. We take advantage of two unique properties of fibers. First, the transmission through the fiber is extremely high, even in the presence of strong mode mixing, thus information on the input state of the light is only scrambled but not lost. This is in contrast to highly scattering (diffusive) samples where most of the incident light is reflected. Second, unlike highly scattering samples, fibers allow to fully control the coupling of the input light to all the guided modes thanks to the finite numerical aperture.

**Results**

Unlike scattering media, in optical fibers backscattering is negligible. However, we can take advantage of the versatility of fibers to study CBS in diverse configurations. We start with the simplest configuration, a double passage configuration, in analogy to double passage through distorting phase screens[18,19]. Due to the Fresnel reflection at the output facet of the fiber, light which propagates through the fiber can be reflected back towards the input facet. The light can then propagate along time-reversed paths (Fig. 1(b)). As in typical CBS experiments with scattering samples, to maximize the CBS enhancement we illuminated the fiber with a well-defined incident angle[20], detected the light at the far-field of the fiber facet, and measured the co-polarization channel (Fig. 1(c), see also Methods). The fiber was a 5 meter long step-index multimode fiber which supports ~1500 guided modes. The guided modes were strongly coupled, due to fiber imperfections and stress induced by bending the fiber. The interference between the guided modes results in a random grainy pattern called speckle that is measured by a CCD camera. After averaging over 200 distinct speckle patterns that were recorded while the fiber was constantly perturbed, a smooth intensity distribution was obtained (Fig. 1(d)). The average intensity exhibits two bright regions: a saturated spot (red bottom arrow) due to the specular reflection from the front facet, and the CBS signal (blue top arrow). The two bright regions are separated because we tilted the fiber facet relative to the angle of input beam. Similar to phase conjugation[21], the CBS is observed exactly in the opposite direction to the input beam, whereas the specular reflection is observed at the mirrored position. The width of the CBS enhancement area is determined by the diffraction limit, i.e. it is equal to the average width of a single speckle grain. It is inversely proportional to the diameter of the fiber core and does not depend on the fiber length.

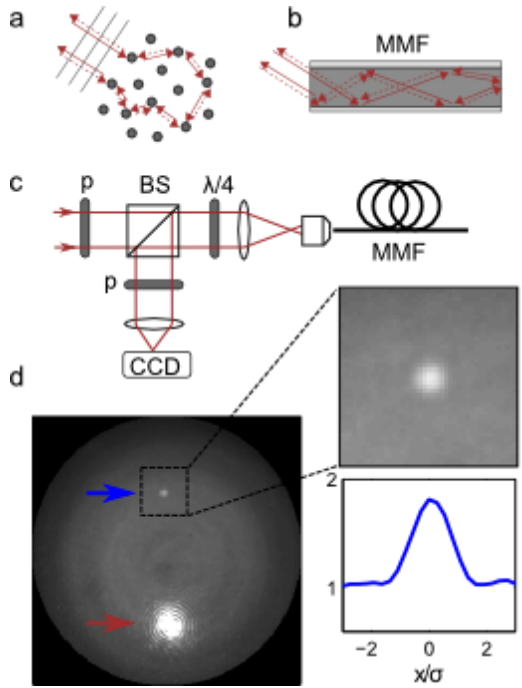

**Figure 1. Coherent backscattering of light in a multimode fiber (MMF).** (a) Illustration of CBS in a random scattering sample, depicting one pair of time-reversed paths (solid and dashed arrows). The two paths accumulate the same phase inside the sample and interfere constructively in the direction opposite to the incident wave. (b) Illustration of CBS in a MMF. Due to fiber imperfections, twisting and bending, the guided modes are strongly coupled, and the light can propagate through many different paths. Part of the light is back reflected from the output end of the fiber due to Fresnel reflection, creating time-revered paths as illustrated. (c) Schematic of the experimental setup for observing CBS in a MMF. A laser beam ($\lambda = 640$ nm) is coupled to a 5 meter long step-index multimode fiber that supports ~1500 guided modes. The incident beam is collimated and its direction deviates from the normal of the input facet of the fiber, so that the backscattering direction differs from that of specular reflection. The back-reflected light is picked up with a beam splitter (BS) and recorded by a CCD camera. An additional lens is used to image the far-field of the fiber front facet to the camera. The light impinging on the fiber is circularly polarized to suppress specular reflection from the front facet of the fiber (P – linear polarizer, $\lambda/4$ – quarter-wave plate). The linear polarizer is placed in front of the camera to detect signal in the same polarization as the input (co-polarization channel). (d) Far-field intensity distribution of the returning light, averaged over 200 fiber configurations. Enhanced intensity in the backscattering direction is observed (top blue arrow), and in its mirrored position is a strong specular reflection (red bottom arrow). The insets show a magnified view of the CBS peak, and a cross section of the averaged intensity, with an enhancement factor of 1.81. $\sigma$ is the full width half maximum of a single speckle grain.

The above example shows that CBS exists in multimode fibers, in a double passage setting that resembles CBS in scattering media. In the following, we studied CBS in a new configuration, which takes advantage of the high transmission through fibers. Specifically we investigated whether the light that is transmitted through the fiber can also exhibit interference between time-reversed paths. To this end, we injected light to both ends of the fiber by splitting the input beam by a beam splitter (BS1 in Fig. 2(a)). The counter-propagating fields inside the fiber were combined by BS1, forming a Sagnac loop. The light can propagate through many different paths inside the fiber, yet for every path that propagates in the clockwise direction, there is a reciprocal path that propagates in the counter-clockwise direction. Every pair of reciprocal paths accumulates the same phase. Thus, when measuring the co-polarization channel, the two counter-propagating fields interfere constructively on a camera which images the light that propagates back towards the source (CCD1), resulting in a CBS peak in Fig. 2(b).

Interestingly, the loop configuration also allows us to observe a destructive interference between time-reversed paths, by recording the light that exits from the second port of BS1 (CCD2). A dip rather, than a peak, was then observed. This, however, does not indicate that reciprocity was broken. Since the illumination and detection were performed at different ports of BS1, strictly speaking this is not a CBS configuration. The mechanism for the destructive interference is the $\pi/2$ phase that is associated with each reflection from a lossless beam splitter. Thus, the counter-clockwise paths accumulate a $\pi$ phase shift relative to the clockwise paths, as they are reflected twice by the beam splitter (BS1). We note that to identify a region of perfect destructive interference requires averaging over just a few speckle realizations. Despite of speckle variation from one realization to another, the destructive interference between the two paths always produces a null intensity at the same position, whereas the locations of other null intensities due to interference of many random paths would vary with realizations, and the probability to detect a null intensity for a sum of even just two uncorrelated speckle patterns is negligible. In contrast, to observe a CBS peak it is necessary to average over many more realizations, since the intensity measured at the peak position fluctuates between realizations.

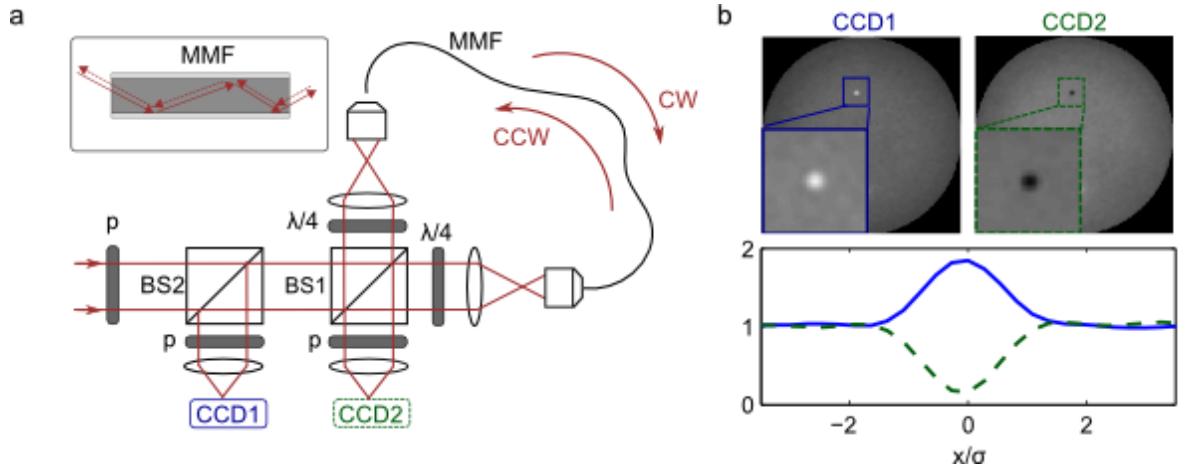

**Figure 2. Coherent backscattering in a multimode fiber (MMF) loop.** (a) A collimated laser beam is split by a beam splitter (BS1) and coupled to the two end facets of a 6 meter long step-index multimode fiber. The transmitted fields are recombined by BS1, forming a Sagnac loop. The two facets of the fiber are positioned at conjugated planes, and their combined far-field images are recorded by CCD1 and CCD2. Inset shows an illustration of one pair of counter-propagating paths in the fiber. (b) Top panel: Images recorded by the two cameras, averaged over 200 configurations of the fiber. On CCD1 a CBS peak is observed due to constructive interference of time-reversed paths, and on CCD2 a CBS dip due to destructive interference. Since the illumination and detection channels are at different ports of BS1, the destructive interference in CCD2 does not violate the reciprocity principle. Bottom panel: intensity distribution in the vicinity of backscattering direction (CCD1: blue solid line, CCD2: green dashed line). The CBS peak to background ratio is 1.85, and the dip to background ratio is 0.15.

Next, we show that it is possible to control CBS, by tuning the relative phase between the time-reversed paths in the multimode fiber. This requires breaking optical reciprocity, which typically results in suppression of CBS[1–3]. The reason is that in non-reciprocal systems, the time-reversed paths accumulate different phases. If the phase difference between each pair of time-reversed paths depends on its trajectory, then the superposition of all the pairs smears out the interference effect and CBS is suppressed. This happened in scattering media with a strong magneto-optical response, where different paths encountered a different overlap with the magnetic field[1]. However, multimode fibers allow the same nonreciprocal phase to be imposed on all pairs of time-reversed paths. We achieved this by adding a Faraday rotator to the Sagnac interferometer, a configuration which was previously considered for optical switches[22]. A Faraday rotator is comprised of a magnetic-optical crystal and a permanent magnet producing a magnetic field that is orientated along the propagation direction. It introduces opposite phase delays for right and left circularly polarized light. The magnitude of the phase delay depends on the strength of the magnetic field, and on the angle between the magnetic field and the propagation direction. Most importantly, reciprocity is broken because the phase delay for light with the same circular polarization has opposite signs when the propagation direction is parallel or anti-parallel to the magnetic field.

We placed the Faraday rotator in between two segments of the multimode fiber (5 meter and 1 meter long), and added a collimating lens on each side of the rotator (Fig. 3(a)). The beam coming out of the fiber was therefore nearly collimated when passing through the rotator, and thus the angle between the magnetic field and the propagation direction for each of the paths through the rotator was approximately the same. This does not mean that all the paths experience the same non-reciprocal phase. The strong mode mixing in our multimode fiber completely scrambles the polarization of the light, thus at the input surface of the Faraday rotator, the polarization state of each path has a different composition of the left and right circular polarization components. Since the two circular polarizations acquire an opposite phase inside the Faraday rotator, the net effect is path-dependent. Nevertheless, the non-reciprocal effect is insensitive to the polarization scrambling and is not washed out upon ensemble averaging. The reason is that for each pair of time-reversed paths, the circular polarization decomposition is identical at the input of the rotator. For example, if the Faraday rotator imposes a phase of $\varphi/2$ on the right circular polarized light propagating in the clockwise direction, it will impose a phase of $-\varphi/2$ on the right circular polarized light propagating in the counter-clockwise direction. Then for each pair of time-reversed paths the phase *difference* between the clockwise and the counterclockwise paths is $\varphi$ for the right circular polarization component, and $-\varphi$ for the left circular polarization component. Hence, regardless of the polarization decomposition of the polarization state at the input of the Faraday rotator, the interference between each pair of time-reversed paths produces the same $\cos(\varphi)$ modulation of backscattered intensity.

By adjusting the distance between the crystal and the permanent magnet, we controlled the effective strength of the magneto-optical effect and tuned the non-reciprocal phase, $\varphi$. Figure 3 depicts the average intensities recorded by the cameras CCD1 and CCD2, for different values of the non-reciprocal phase, $\varphi$, applied by the Faraday rotator. The CBS peak on CCD1 (with the peak to background ratio of 1.86) turns into a dip (with the dip to background ratio of 0.15) at $\varphi=\pi$. Similarly, the dip on CCD2 turns into a peak. This transition demonstrates that the interference between reciprocal paths inside the fiber can be continuously tuned from constructive to destructive with a visibility of 85%. Moreover, this behavior survives the strong mode mixing in the multimode fiber, and can therefore be used to enable robust control of CBS.

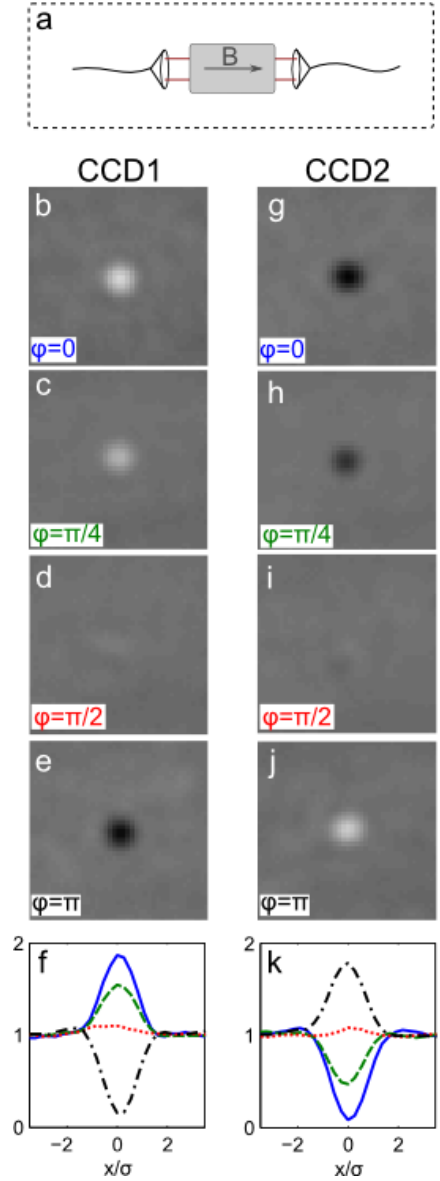

**Figure 3. Control of coherent backscattering with a Faraday rotator.** (a) A Faraday rotator is inserted to the fiber loop shown in Fig. 2, by splitting the multimode fiber to two segments (5 mater and 1 meter long) and placing the Faraday rotator in between. Two collimators are placed on both sides of the Faraday rotator to collimate the beams coming out of the fiber and refocusing them into the other fiber. (b)-(e) Magnified view of the CBS signal measured by CCD1, for four nonreciprocal phases φ induced by the Faraday rotator. φ = 0 (b), π/4 (c), π/2 (d), and π (e). A continuous transition is observed from a CBS peak (peak to background ratio of 1.86) to a dip (dip to background ratio of 0.15). (f) Cross sections of the images (b)-(e) (φ = 0 blue solid line, φ = π/4 green dashed line, φ = π/2 red dotted line, and φ = π black dashed-dotted line). (g)-(k): same as (b)-(f) for the images recorded by CCD2.

**Discussion**

In this work, we demonstrated two mechanisms for observing robust destructive interference between pairs of counter-propagating paths inside the multimode fiber. In the first demonstration, the destructive interference was possible without breaking reciprocity, because the extra port of the beam splitter provided access to returning light in a final state that is orthogonal to the input state. Similarly, weak anti-localization without breaking reciprocity was observed for electrons in thin metallic films[10,23], and in quantum dots[24], with strong spin-orbit coupling which causes the spin of the output state to be antiparallel to the spin of the input state. A related effect was observed in graphene, where rotation of the pseudospin along the propagation path also results in weak anti-localization[25]. In optics, it was predicted that in photonic graphene lattices, the pseudospin of the backscattered wave is anti-parallel to the input pseudospin direction, resulting in a CBS dip due to a Berry phase of $\pi$[26]. Consequently, enhanced transmission in photonic graphene lattices was observed in the microwave regime[27]. Here, we directly measured the destructive interference between the counter-propagating paths inside the fiber.

In the second approach, we broke reciprocity in order to observe the CBS dip, i.e. the destructive interference was between strictly time-reversed paths. Reciprocity breaking mechanisms for obtaining a CBS dip were only theoretically proposed before, e.g. for scattering of light from ultracold atoms, where reciprocity was broken either by the magnetic field and Zeeman splitting[28], or by nonlinear light-matter interactions[29,30]. Our approach using a magneto-optical effect, allows not only a direct observation of the CBS dip, but also a precise control of the relative phase between the time-reversed paths. We demonstrated a continuous transition from a CBS peak to a dip, with a visibility that is orders of magnitudes larger than the visibility of the oscillations in magnetoresistance measurements of electrons subject to a non-reciprocal Aharonov-Bohm phase[11,12].

In conclusion, we have shown that multimode fibers can serve as versatile platform for studying fundamental physical phenomena in complex systems. In particular, we developed a new configuration for manipulating CBS and demonstrated a precise control that is robust against external perturbations. Multimode fibers further provide exceptional opportunities for exploring fundamental aspects of complex coherent phenomena, e.g., by increasing the complexity of the system through optical nonlinearities or by using chaotic fibers[31]. These aspects are expected to have practical implications to imaging and communication applications that are based on multimode fibers.

## Methods

*CBS setup.* We spatially filtered a linearly polarized CW laser beam (λ = 640nm, ORBIS LX, Coherent), and clipped it with an iris to create a nearly flat-top beam. We then coupled the beam to the multimode fiber by demagnifying it using a lens and a microscope objective (x20, NA=0.4). The beam at the input facet slightly overfilled the core of the fiber (diameter D=105um). To separate the specular reflection from the CBS signal we tilted the facet of the fiber. The returning light was collected with the same objective, and the back focal plane of the objective was reimaged onto the CCD camera, i.e. we recorded the Fourier plane of the fiber facet. We used a linear polarizer in front of the CCD camera to measure the same polarization as the incident beam (i.e. co-polarization channel). We placed a quarter-wave plate before the fiber, orientated at 45 degrees relative to the incident polarization direction, in order to reduce the specular reflection from the input facet of the fiber. The above setting was used for the double passage and the Sagnac configurations.

*Multimode fiber.* In all the configurations we studied, we used a standard step-index multimode fiber, with a numerical aperture of 0.22, and a core diameter of 105um. We twisted and bended the fiber to enhance mode mixing. All our measurements were performed in the regime of strong mode mixing, which created a homogenous distribution of the ensemble-averaged intensity at the far-field of the fiber facet.

## Acknowledgements
We thank Doug Stone, Dan Prober and Shanhui Fan for fruitful discussions. We acknowledge support from the National Science Foundation under Grant No. DMR-1205307 and the Office of Naval Research under Grant No. ONR MURI SP0001135605.